\documentclass[a4paper]{jpconf}
\usepackage{graphicx}
\begin{document}
\title{The CUORE slow monitoring systems}

\author{L Gladstone$^1$, 
D~Biare$^2$, 
L~Cappelli$^3$, 
J~S~Cushman$^4$, 
F~Del~Corso$^5$, 
B~K~Fujikawa$^2$, 
K~P~Hickerson$^6$, 
N~Moggi$^{5,7}$, 
C~E~Pagliarone$^8$, 
B~Schmidt$^2$, 
S~L~Wagaarachchi$^{2,9}$, 
B~Welliver$^2$ and 
L~A~Winslow$^1$ 
} 

\address{$^1$ Massachusetts Institute of Technology, Cambridge, MA 02139, USA}
\address{$^2$ Nuclear Science Division, Lawrence Berkeley National Laboratory, Berkeley, CA 94720, USA}
\address{$^3$ INFN -- Sezione di Genova, Genova I-16146, Italy}
\address{$^4$ Department of Physics, Yale University, New Haven, CT 06520, USA}
\address{$^5$ INFN Sezione di Bologna, Bologna I-40127, Italy}
\address{$^6$ Department of Physics and Astronomy, University of California, Los Angeles, CA 90095, USA}
\address{$^7$ Dipartimento di Scienze per la Qualita’ della Vita, Alma Mater Studiorum Universita’ di Bologna, Bologna I-47921, Italy}
\address{$^8$ Dipartimento di Ingegneria Civile e Meccanica, Universit\`{a} degli Studi di Cassino e del Lazio Meridionale, Cassino I-03043, Italy}
\address{$^9$ Department of Physics, University of California, Berkeley, CA 94720, USA}

\ead{gladston@mit.edu}

\begin{abstract}
CUORE is a cryogenic experiment searching primarily for neutrinoless double decay in $^{130}$Te. It will begin data-taking operations in 2016. To monitor the cryostat and detector during commissioning and data taking, we have designed and developed Slow Monitoring systems. In addition to real-time systems using LabVIEW, we have an alarm, analysis, and archiving website that uses MongoDB, AngularJS, and Bootstrap software. These modern, state of the art software packages make the monitoring system transparent, easily maintainable, and accessible on many platforms including mobile devices.
\end{abstract}

\section{The CUORE Cryostat and Detector}

The Cryogenic Underground Observatory for Rare Events (CUORE)~\cite{Detector} is an experiment at Laboratori Nationali di Gran Sasso (LNGS) in central Italy. CUORE searches primarily for neutrinoless double beta decay. The detector consists of 988 TeO$_2$ crystals housed in a large~\cite{ColdestMeter} helium dilution cryostat that was custom-made for CUORE and operates below 10\,mK~\cite{Cryostat,Stefano}. 
The cryostat contains 6 interior vessels, each progressively colder. 
Building, cooling, and running CUORE requires a complex interplay of systems that need to be monitored. 

\section{LabVIEW Monitoring and Control}

Several of the subsystems are controlled by National Instruments LabVIEW. The programs for control and logging are a mix of custom-made software and proprietary drivers. 

Some of the systems include: the temperature at each layer of shielding while the cryostat cools; the flow of nitrogen flushing that abates radon buildup around the detector crystals during storage and installation; the pressures and temperatures from the pulse tube cryocoolers; and the positions, pressures, and loads from the calibration system. Other systems can be added fairly simply. Some systems included in the Slow Monitoring don't use LabVIEW at all (e.g. monitoring the temperature from the readout electronics racks), but still provide text-based log files.

Because of the variety of systems involved, the log files are necessarily different from each other. An unavoidable difference is that some systems log data more frequently than others, ranging from each second to every 5 minutes, with gaps that can last several months between datapoints. Another difference between the systems is the format of the log files. While the format varies between systems, they are all text-based, contain a timestamp accurate to within a second, and contain a new line for each new timestamp. Most are tab-delimited with text column headers. The date and time fields are particularly varried, including an extreme case where, instead of logging the date and time for each measurement, the controller logs only the time difference between the datapoint and the start of the file. The date-time format for each system is treated as a special case when the log files are later parsed.


\section{Slow Monitoring Dataflow}
The Slow Monitoring systems collect the variety of log files, archive them, and display them on a website using the following dataflow. 

First, within the protected lab network, each LabVIEW controller writes a text-based log file, as described above. A single server processes these files. The server runs a cron script each minute, which checks each subsystem computer for new log files and copies them to the central server. Any new files (based on their timestamps) are processed line-by-line and written to a central database. 

The database uses MongoDB, an unschemed, well-documented database protocol that interfaces well with Python scripts. The Slow Monitoring database is split into collections by month, with additional collections for metadata and website user account information. A single MongoDB instance houses both the Slow Monitoring database described here and the CORC database mentioned below. The databases are backed up using a standard Mongo replication scheme with additional servers both at LNGS and off-site. 

The basic design and configuration of this system comes from within the CUORE collaboration, and some key aspects of the Slow Monitoring were implemented by the Parthanon Software Group~\cite{Parthanon}. 

\section{Slow Monitoring Plotting Tools}

The data are displayed from the database on a website accessible to the CUORE collaboration. 
This website is based on the AngularJS and Bootstrap libraries, with jQuery for database calls and HighCharts for plotting. These libraries make the page layout accessible to various screen sizes, including tablets and smartphones. This allows the collaboration to monitor CUORE's status off-site, for example from a commuter train. The Slow Monitoring part of the webpage is the most flexible part of the central monitoring systems, and is most relevant to the collaboration during changes in the detector configuration. 

The Slow Monitoring website has several parts. There are two main plot interfaces: an updating ``overview'' page and a fully-interactive ``detector'' page. 

The overview page contains a few plots of key parameters over the last few hours. The choice of parameters and time periods is configurable by website administrators, and varies with time. For example, during the cryostat cooling period, the most relevant parameters may be the thermometers currently in their most sensitive range. While the cryostat is open and the detector crystals are exposed, the relevant parameters would include the ambient radon levels. These plots query the database every 10 seconds for updates. 

The detector page allows users to plot any variable stored in the database. Up to about 10 variables can be plotted at once. The database queries are split up by the plot bins, returning about 1- to 2 thousand points total per variable. If the data is fairly uniform in time, this makes the database queries much faster. If the data is not uniform, the query is still never slower than simply returning all the data. The plots can be downloaded as images, and the datapoints can be exported as text files. When website users make plots, the database query returns data within about a second for each request.

\section{CORC Plotting Tools}

The CUORE Online Run Control (CORC) system works from the same database and website as Slow Monitoring. It displays the central data from CUORE: the response of the detector bolometers. It shows the rates from each detector channel, in both summary form and with pop-up plots for individual detector crystals. There are intricate controls to change the display format, global color scheme and range, and channel selection. CORC provides the interface for the early stages of data quality monitoring and analysis: analysis shifters can manually tag intervals of bad or questionable data. Some bad intervals are tagged automatically, ensuring a standard baseline tagging algorithm. Complementing Slow Monitoring, the CORC site is most relevant to the collaboration during steady data-taking. 

\section{Alarms}

As data is put into the database, it goes through an alarm system. Alarms can be configured both for groups, like shifters or cryogenics experts, and for individuals. The alarm configurations are stored in the database. When new data is imported, it is checked against the list of configured alarms. If the data triggers any alarms, the trigger instance is displayed on the website. An email goes to the user or group who configured the alarm. An advanced feature is currently being debugged which calls users' telephones. This alarm system is an integral part of the detector monitoring shifts.

The ``alarm delay'' is measured between the alarm-triggering data being collected and an alarm reaching a user, expressing the total system processing time. The goal is to limit this time to about 2 minutes average, dominated by the 1 minute between instances of running the Slow Monitoring cron job.

\section{Nagios Server Monitoring}

All of the servers mentioned above are monitored using Nagios. This includes 18 ``hosts'' and 159 ``services''. The services include the processes to host websites and databases. If any host or service loses connectivity, Nagios sends persistant emails until the problem is acknowledged. The log files from Nagios can be included in Slow Monitoring.

\section{Conclusion}
CUORE has a flexible system in place to monitor data from the cryostat and detector both for the general state of the experiment and for physics analysis. All of these monitoring systems have run during the detector installation this year, and will continue to run as CUORE starts taking data. Access to data online and on mobile devices means shifters can monitor from off-site. Most importantly, the ease of this interface allows collaborators to focus on the physics represented in the data.

\end{document}